\documentclass[12pt]{article}
\usepackage{amsfonts}

\font\mybb=msbm10 at 12pt
\def\bb#1{\hbox{\mybb#1}}

\begin{document}
\thispagestyle{empty}

\begin{flushright}
January, 2010
\end{flushright}

\vskip 2.5cm

\begin{center}
\baselineskip=16pt {\LARGE\bf  Superembedding approach to M$0$-brane and multiple M$0$--brane system}

\vskip 1.5cm

 {\large\bf Igor A. Bandos\footnote{Also at Institute for Theoretical Physcs,
NSC Kharkov Institute of Physics \& Technology,
UA 61108,  Kharkov, Ukraine. E-mail: igor\_bandos@ehu.es, bandos@ific.uv.es }}
\vskip 1.0cm {\it\small $^\dagger$ IKERBASQUE, the Basque Foundation for Science, and  \\ Department of Theoretical Physics,
University of the Basque Country,  \\ P.O. Box 644, 48080 Bilbao,
Spain
}
  \vspace{24pt}

\end{center}

\vskip 0.5cm

\par
\begin{quote}

We study the possibility to describe multiple M0--brane system in the frame of superembedding approach. The simplest framework is provided by the maximally supersymmetric non-Abelian $SU(N)$ Yang--Mills  supermultiplet on the  $d=1$ ${\cal N}=16$  superspace the embedding of which to the target $D=11$ supergravity superspace is determined by the so-called superembedding equation, characteristic of the worldline superspace of a single M$0$-brane. We use it to obtain a covariant generalization of the Matrix model  equations describing the multiple M$0$--system  in flat target superspace.

\end{quote}

\newpage

\setcounter{page}{2}
\section{Introduction}
\setcounter{equation}{0}

The action for M$0$-brane, the Brink-Schwarz action for massless eleven dimensional superparticle, was discussed for the first time by Bergshoeff and Townsend in \cite{B+T=Dpac}, were it was used as a starting point to obtain the 10D type IIA D$0$-brane action, the $p=0$ representative of the D$p$-branes (Dirichlet $p$--branes) \cite{Sagnotti,Horava,Polchinski89}.

Although the r\^{o}le of D$p$-branes  and multiple  D$p$-brane system in String/M-theory \cite{M-theory}  was appreciated in middle 90th \cite{Polchinski95}, the problem of Lorentz covariant and supersymmetric action which governs  dynamics of the system of $N$ D$p$-branes still remains unsolved. Even before the nonlinear action
for single D$p$--brane were found in \cite{Townsend95,Dpac,Dpac2,B+T=Dpac}, it was understood that at the very low energy approximation this should reduce to  the action of maximally supersymmetric Yang--Mills theory (SYM) \cite{Witten:1995im,D0-1996}, which can be obtained by dimensional reduction of $D=10$ SYM down to $d=p+1$. The purely bosonic action for multiple D$p$-brane system was proposed by Myers \cite{Myers:1999ps}; he derived it by chain of dualities starting from the 10D non-Abelian DBI--action with symmetric trace prescription proposed by Tseytlin \cite{Tseytlin:DBInA}. However, more than decade of attempts to make supersymmetric and Lorentz covariant generalization of the action \cite{Myers:1999ps} and/or \cite{Tseytlin:DBInA} was not successful (although a particular progress was achieved in the cases of low dimensions $D$, low dimensional branes  and low co-dimensional branes
\cite{Dima01,Drummond:2002kg}).

The only exception is the  boundary fermion approach by Howe, Lindstr\"om and Wulff  \cite{Howe+Linstrom+Linus,Howe+Linstrom+Linus=2007} which does provide supersymmetric and covariant description of Dirichlet branes, but on the 'pure classical' (or 'menus one quantization') level in the sense that to arrive at the description of multiple D-brane system in terms of the variables corresponding to the standard single D$p$--brane action \cite{Dpac,Dpac2,B+T=Dpac} (usually considered as a classical or quasi-classical action) one has to perform a quantization of the boundary fermion sector. The complete quantization of the model \cite{Howe+Linstrom+Linus,Howe+Linstrom+Linus=2007} should produce not only worldvolume fields of multiple D$p$-brane system but also bulk supergravity and higher stringy modes. The partial quantization of only the boundary fermion sector  allowed Howe, Lindstr\"om and Wulff to reproduce the purely bosonic Myers action \cite{Myers:1999ps}, but the Lorentz invariance was lost on this way. The supersymmetric action \cite{Howe+Linstrom+Linus=2007}, involving quite interesting and novel prescription of  integration,  possesses the $\kappa$--symmetry with parameters dependent of boundary fermions. This implies that, upon the quantization, the action with non-Abelian $\kappa$--symmetry should appear.  The previous attempts to incorporate non-Abelian  $\kappa$--symmetry gave negative results \cite{Bergshoeff:2001dc}.
Probably, the further development of the boundary fermion approach will help, in particular, to resolve this problem as well as other problems which hamper the way to constructing the supersymmetric and Lorentz covariant (diffeomorphism invariant) action for multiple D$p$--brane system similar to the DBI plus WZ action for single D$p$--brane  \cite{Dpac,Dpac2,B+T=Dpac} \footnote{DBI is for Dirac--Born--Infeld and WZ is for Wess--Zumino. The action of \cite{Dpac,Dpac2,B+T=Dpac} is given by the sum of Dirac--Born--Infeld, kinetic term  and of the Wess--Zumino term describing coupling to the RR gauge fields, the antisymmetric tensor (differential form) potentials $C_{p+1}$, $C_{p-1}$, $\ldots\;$.}. It also might happen that the boundary fermion action of \cite{Howe+Linstrom+Linus=2007} is the best what one can write to describe multiple D$p$--brane system without involving additional fields carrying the bulk  supergravity and/or  other stringy degrees of freedom. In both cases in the middle time it looks reasonable, not diminishing the importance of the development of boundary fermion approach, to search for a more straightforward, probably approximate but covariant and supersymmetric description, going beyond the $U(N)$ SYM approximation, and formulated in variables  similar to the ones used to write  single D$p$--brane action \cite{Dpac,Dpac2,B+T=Dpac}.

Recently \cite{IB09:D0} a way to search for  multiple D$p$-brane equations in the frame of superembedding approach  \cite{bpstv,hs96,hs2,bst97,HS+Chu=PLB98,Dima99,IB09:M-D} was proposed and elaborated for the case of D$0$-branes. Roughly speaking the proposition of \cite{IB09:D0} is to describe the $N$ D$p$--brane system by imposing a suitable set of constraints on field strengths of the maximal $d=p+1$ dimensional $SU(N)$ SYM super-1-form potential which lives on the worldvolume superspace ${\cal W}^{(p+1|16)}$ with $d=p+1$ bosonic and $16$ fermionic directions the embedding of which into the type II supergravity superspace is determined by the so--called superembedding equation. The latter universal equation describes completely the dynamics of single D$p$-brane with $p<6$ \cite{HS+Chu=PLB98}.

The selfconsistency of the above system of equations has been checked presently for the case of multiple D$0$--brane (multiple D-particle) system. It was shown \cite{IB09:D0} that the resulting system of equations describes the famous dielectric brane effect of Emparan and Myers \cite{Emparan:1997rt,Myers:1999ps} which consists in such a polarization of the multiple D-brane system by external fluxes that results in interaction with higher form gauge fields ($C_q$ with order $q>p+1$) which do not interact with a single D$p$-brane. In the case of flat target superspace our multiple D$0$--brane equations coincide with the result of dimensional reduction of 10D SYM equation to $d=1$ \cite{IB09:D0}. This is important because it shows the relation with Matrix model \cite{Banks:1996vh} the description of which uses the $d=1$ reduction of maximal SYM action  as a  starting point.

However, one of the most important property of the  Matrix model is the 11D invariance. The possibility of the restoration of the 11D Lorentz invariance in the system of 10D D$0$-branes as described by 1d reduction of the 10D SYM action, was one of the key points in \cite{Banks:1996vh}  which allowed the M(atrix) model conjecture on that the Matrix model considered as multiple D$0$-brane system provides a nonperturbative description of the underlying String/M-theory. In this respect the natural question is whether one can show the restoration of 11D Lorentz invariance in the framework of superembedding description of multiple D$0$-brane proposed in \cite{IB09:D0}.

It seems that the simplest way to check this is to try to develop the similar description of 11D multiple M$0$--brane system. Indeed, as far as D$0$--brane can be obtained by `dimensional reduction'/duality of  M$0$-brane \cite{B+T=Dpac}, it is natural to expect that multiple D$0$-brane system can be obtained by  `dimensional reduction'/dualization of the system of multiple M$0$-branes, if such exists. Then one can also expect  that, as in the case of single branes, this `dimensional reduction'/dualization is invertible, and the inverse transformation would give the desired restoration of the 11D Lorentz invariance in the multiple D$0$-brane system described by superembedding approach of \cite{IB09:D0}.


Furthermore, the possible existence of multiple M$0$-brane system may be interesting on its own.
The fact that, in distinction to D-particles (D$0$-branes), M-particles (M$0$-s) are massless, suggests a possibility of appearance of novel features and/or novel problems when studying multiple M$0$-system.
The aim of the presence letter is to check whether one can develop a description of such a system in the frame of superembedding approach on the line of \cite{IB09:D0}.

We begin in the next Sec. 2 by  a brief review of superembedding approach on the example of single M$0$-brane. This allows us to fix the notation and to provide the basis for the superembedding description of multiple M$0$ system which we develop in Sec. 3. The concluding Sec. 4 contains discussion and outlook.

\section{Superembedding approach to a single M$0$-brane.  }
\setcounter{equation}{0}

\subsection{Basic notation. Target $D=11$ superspace $\Sigma^{(11|32)}$}

We denote the supervielbein of target  D=11  superspace $\Sigma^{(11|32)}$ by
\begin{eqnarray}\label{Eua-cE}
{E}^{ {A}}:=
dZ^{ {M}}E_{ {M}}{}^{ {A}}(Z)=
 ({E}^{ {a}}, {\cal E}^{ {\alpha}})\; , \qquad \cases{ { {\alpha}}=1,\ldots ,
32 \; , \cr { {a}}=0,1,\ldots, 9, 10 \; } \; .   \qquad
\end{eqnarray}
and its local coordinates by
\begin{eqnarray}\label{Z=x,th}
Z^{{ {M}}}= (x^{ {m}}\, ,
\theta^{\check{ {\alpha}}})\; , \quad
{\check{ {\alpha}}}=1,\ldots , 32\; , \qquad { {m}}=0,1,\ldots, 9, 10   \; . \qquad
\end{eqnarray}
The supervielbein (\ref{Eua-cE}) describes 11D supergravity when it obeys the set of superspace constraints
\cite{CremmerFerrara80BrinkHowe80} the most essential of which are collected in the following expression for the
bosonic torsion two form:
\begin{eqnarray}
\label{Ta=11D} & T^{ {a}}:= DE^{ {a}} =
-i{\cal E}\wedge \Gamma^{ {a}} {\cal E} \; . \qquad
\end{eqnarray}
In (\ref{Ta=11D}) ${\cal E}\wedge \Gamma^{ {a}} {\cal E}:= {\cal E}^{ {\alpha}}\wedge
{\Gamma}^{ {a}}_{ {\alpha} {\beta}}  {\cal E}^{ {\beta}}$, \footnote{Here and below we write explicitly the exterior product symbol $\wedge$. The exterior product of a $q$-form $\Omega_q$ and a $p$-form $\Omega_p$ has the property $\Omega_q \wedge \Omega_p= (-1)^{p q} \Omega_p \wedge \Omega_q $ if at least one of two differential forms is bosonic; when both are fermionic, an additional $(-1)$ multiplier appears in the {\it r.h.s.}. The exterior derivative acts on the products of the forms `from the right': $d(\Omega_q \wedge \Omega_p)=  \Omega_q \wedge d\Omega_p + (-1)^{p} d\Omega_q \wedge \Omega_p$. }
the symmetric matrices  ${\Gamma}^{ {a}}_{ {\alpha} {\beta}}  = ({\Gamma}^{ {a}}C)_{ {\alpha} {\beta}}= {\Gamma}^{ {a}}_{ {\beta} {\alpha}}$ are real,  while  the 11D Dirac matrices $({\Gamma}^{ {a}}){}_{ {\alpha}}{}^{ {\beta}}$ and charge conjugation matrix  $C$  are imaginary in our mostly minus notation $
\eta^{ {a} {b}}= diag(+,- , \ldots, -)$. Below we will mostly use  ${\Gamma}^{ {a}}_{ {\alpha} {\beta}}$ and their counterparts with upper case  indices, the real matrices $\tilde{\Gamma}_{ {a}}{}^{ {\alpha} {\beta}}  = ({\Gamma}_{ {a}}C)^{ {\alpha} {\beta}}= \tilde{\Gamma}_{ {a}}{}^{ {\beta} {\alpha}}$.

Studying Bianchi identities with (\ref{Ta=11D}) one finds the structure of fermionic torsion and $SO(1,10)$ curvature two form
\cite{CremmerFerrara80BrinkHowe80}
\begin{eqnarray} \label{Tf=}
T^\alpha&=&- {i \over 18} E^a \wedge E^\beta
\left(F_{ac_1c_2c_3}\Gamma^{c_1c_2c_3}{}_\beta^{\;\alpha} + {1\over 8}
F^{c_1c_2c_3c_4} \Gamma_{a c_1c_2c_3c_4}{}_\beta^{\;\alpha}\right) +
{1 \over 2} E^a \wedge E^b T_{ba}{}^\alpha(Z) \; , \qquad \\
\label{RL=} R^{ab}&=&E^\alpha \wedge E^\beta \left( -{1\over3}
F^{abc_1c_2}\Gamma_{c_1c_2} + {i\over 3^. 5!} (\ast
F)^{abc_1\ldots c_5} \Gamma_{c_1\ldots c_5} \right)_{\alpha\beta}
+ \nonumber
\\ &&   \qquad +  E^c \wedge E^\alpha \left(
-iT^{ab\beta}\Gamma_{c}{}_{\beta\alpha} + 2i T_c{}^{[a \, \beta}
\Gamma^{b]}{}_{\beta\alpha} \right) + {1 \over 2} E^d \wedge E^c
R_{cd}{}^{ab}(Z) \; .
\end{eqnarray}
Here $F_{abcd}= F_{[abcd]}(Z)$ is the bosonic superfield strength of the 3-form gauge field of the 11D supergravity. It obeys $D_{[a}F_{bcde]}=0 $ and  $D_{\alpha}F_{abcd}=- 3! T_{[ab}{}^\beta \Gamma_{cd]}{}_{\beta\alpha}$.

 \subsection{Superembedding approach to  M$0$--brane}

\subsubsection{Worldvolume superspaces $W^{(p+1|16)}$ of M-branes}

The standard formulation of M--branes (M$p$-branes  with $p=0,2,5$)  deals with embedding of a purely bosonic worldvolume ${W}^{p+1}$ (worldline for M$0$-case of \cite{B+T=Dpac}) into the target superspace $\Sigma^{(11|32)}$. The superembedding approach to M-branes \cite{bpstv,hs2}, following the so--called STV (Sorokin--Tkach--Volkov) approach to superparticles and
superstrings \cite{stv} (see \cite{Dima99} for review and further references) describes their dynamics in terms of embedding of  {\it worldvolume
superspace} ${\cal W}^{(p+1|16)}$ with $d=p+1$ bosonic and $16$ fermionic directions into the {\it target superspace} $\Sigma^{(11|32)}$.

The embedding can be described in terms of coordinate functions
$\hat{Z}^{{ {M}}}(\zeta)=
(\hat{x}{}^{ {m}}(\zeta)\, ,
\hat{\theta}^{\check{ {\alpha}}}(\zeta))$, which are superfields depending on the local coordinates of  ${\cal W }^{(p+1|16)}$ ($\zeta^{{\cal M}}=(\tau ,\eta^{\check{q}})$,
 with $\check{q}=1,\ldots ,16$ in the case of M$0$-brane)
\begin{eqnarray}
\label{WinS}  {\cal W }^{(p+1|16)}\in \Sigma^{(D|32)} : \hspace{1.5cm}
\fbox{$Z^{ {M}}= \hat{Z}^{ {M}}(\zeta) $}= (\hat{x}^{ {m}}(\zeta)\;
, \hat{\theta}^{\check{ {\alpha}}}(\zeta ))\;  , \qquad
\end{eqnarray}
$ {m}=0,1,.., 9,10 $, $\quad\check{ {\alpha}}=1,...,32$.

To describe M-branes these coordinate functions must obey the {\it superembedding equation} which  completely determines the M$p$-brane dynamics.

\subsubsection{Superembedding equation for M$0$--brane}

To write superembedding  equation  for the case of M$0$--brane, let us denote the supervielbein and local coordinates of the corresponding worldvolume superspace  $W^{1|16)}$ by
\begin{eqnarray}
\label{eA=e0eq} e^A= d\zeta^{{\cal M}} e_{{\cal M}}{}^{A}(\zeta) =
(e^{\# }\; , \; e^{+q}) \; , \qquad  \zeta^{{\cal M}}=(\tau ,\eta^{\check{q}}) \; ,  \qquad \cases{ q=1,\ldots , 16 \; , \cr \check{q}=1,\ldots ,16 \; , } \qquad
\end{eqnarray}
where $q$ is a spinor index of $SO(9)$, $+$ denotes the weight (`charge') with respect to the local $SO(1,1)$ group and  the bosonic vielbein has the weight two, $e^{\# }:=e^{++}$.  Notice that in our notation the upper plus sign is equivalent to the lower minus sign and vice-versa, so that one can equivalently write, for instance, $e^{+q}= e_-^q$ and $e^{\#}=e_{=}:=e_{--}$.

Now, let us denote the pull--back of supervielbein ${E}^{ {A}}(Z)$ of target  superspace, Eq. (\ref{Eua-cE}), to ${\cal W}^{1|16)}$ by  $\hat{E}^{ {A}}:= {E}^{ {A}}(\hat{Z})$. The general form  of its decomposition of the basis of worldvolume supervielbein reads
\begin{eqnarray}
\label{hEa=b+f}
 \hat{E}^{ {A}}:= E^{ {A}}(\hat{Z})=
d\hat{Z}^{ {M}}
E_{ {M}}{}^{ {A}}(\hat{Z}) = e^{\# } \hat{E}_{\# }{}^{
 {A}} + e^{+q} \hat{E}_{+q}{}^{ {A}} \; .
\qquad
\end{eqnarray}
The superembedding equation states that the pull--back of bosonic supervielbein to the worldvolume superspace has no projection on the fermionic supervielbein, {\it i.e.} it reads
\begin{eqnarray}
\label{SembEq=M0}
  \hat{E}_{+q}{}^{ {a}} = 0\; .
\qquad
\end{eqnarray}
After some algebra, one can show that the equivalent form of the superembedding equation (\ref{SembEq=M0}) is given by \cite{bpstv,IB+AN=CQG97}
\begin{eqnarray}
\label{Ei=0}
  \cases{  \hat{E}^{ {a}}  u_{ {a}}{}^i = 0\; , \cr \hat{E}^{ {a}}  u_{ {a}}^{=} = 0\; , }
\qquad
\end{eqnarray}
and also by
\begin{eqnarray}
\label{Eua=e++u--}
  \hat{E}^{ {a}} = {1\over 2} e^{\# } u^{ {a}=} \; ,
\qquad
\end{eqnarray}
where  $u_{ {a}}{}^{\#}=u_{ {a}}{}^{\#}(\zeta)$, $u_{ {a}}{}^{=}=u_{ {a}}{}^{--}(\zeta)$ and $u_{ {a}}{}^i=u_{ {a}}{}^i(\zeta)$ ($i=1,\ldots , 9$) are moving frame vector superfields. These are elements of pseudo--orthogonal moving frame matrix
\begin{eqnarray}
\label{M0:UinSO}
  {U}_{ {a}}^{( {b})}= \left( {1\over 2}( u_{ {a}}{}^{\# }+ u_{ {a}}{}^{=}),  \; u_{ {a}}{}^i\; ,  {1\over 2}( u_{ {a}}{}^{\# }- u_{ {a}}{}^{=})\right)\quad  \in  \quad SO(1,10)  \; ,
\qquad
\end{eqnarray}
which is to say $u_{ {a}}{}^{=}(\zeta)=u_{ {a}}{}^{--}(\zeta)$  and $u_{ {a}}{}^{\#}(\zeta)=u_{ {a}}{}^{++}(\zeta)$ are light--like 11-vectors with zero weight contraction $u_{ {a}}{}^{=}u^{ {a}\# }$ normalized to be equal to $2$, and $u_{ {a}}{}^i(\zeta)$ are 9 orthogonal and normalized  11--vectors superfields which are orthogonal also to $u_{ {a}}{}^{\pm\pm}(\zeta)$,
\begin{eqnarray}\label{u++u++=0}
u_{ {a}}^{\# } u^{ {a}\; \# }=0\; , \qquad  u_{ {a}}^{=} u^{ {a}\; =}=0\; , \qquad \\ \label{u++u--=2} u_{ {a}}^{\; \# } u^{ {a}\; =}=2\; , \qquad   \\  \label{uiuj=-} u_{ {a}}^{\#} u^{ {a}\; i}=0\; , \qquad u_{ {a}}^{=} u^{ {a}\; i}=0\; , \qquad
   u_{ {a}}^{\; i} u^{ {a}\; j}=-\delta^{ij}\; . \qquad
\end{eqnarray}

\subsubsection{Worldline superspace geometry induced by superembedding and moving frame superfields}

Using (\ref{u++u--=2}) one finds that Eqs. (\ref{Eua=e++u--}) contains, in addition to (\ref{Ei=0}), also the conventional constraint which defines the bosonic supervielbein of the worldline superspace  ${\cal W}^{(1|16)}$ to be induced by (super)embedding,
\begin{eqnarray}\label{e++=Eu++}
e^{\# } = \hat{E}^{\# }:= \hat{E}^{ {b}} u_{ {b}} ^{\# }
\; .  \qquad
\end{eqnarray}
To specify further the geometry of ${\cal W}^{(1|16)}$ one has to define as well the $SO(1,1)\otimes SO(9)$ connection and the fermionic supervielbein $e^{+q}$ induced by superembedding\footnote{\label{footK9} The true group of gauge invariance is the small group of massless superparticle, which is the so-called Borel subgroup $[SO(1,1)\otimes SO(9)]\subset\!\!\!\!\!\!\times K_9$ of the Lorentz group $SO(1,10)$, where the $K_9$ transformations are defined by
$\delta u^{\# }= 2k^{\#  i}u^{i}$, $\delta u^{i}= k^{\#  i}u^{=}$
(see \cite{IB07:M0} and refs. therein for details). However, when developing superembedding approach and twistor-like formulations for massless superparticles \cite{IB+AN=CQG97,IB07:M0},  it is convenient to use the formalism where only the gauge symmetry under $[SO(1,1)\otimes SO(9)]$ subgroup of $[SO(1,1)\otimes SO(9)]\subset\!\!\!\!\!\!\times K_9$ is manifest.}.  It is convenient to define this connection by the equations specifying the action of the (exterior)  $SO(1,10)\otimes  SO(1,1)\otimes SO(9)$ covariant derivative on the moving frame vectors
\begin{eqnarray}\label{M0:Du=Om}
Du_{ {b}}{}^{=} = u_{ {b}}{}^i \Omega^{= i}\; , \qquad
Du_{ {b}}{}^{\#} = u_{ {b}}{}^i \Omega^{\# i}\; ,
\qquad   Du_{ {b}}{}^i = {1\over 2} \, u_{ {b}}{}^{\# } \Omega^{=i}+ {1\over 2} \, u_{ {b}}{}^{=} \Omega^{\# i}\; .
\qquad
\end{eqnarray}
These equations involve the generalization of the ${SO(1,10)\over {SO(1,1)\otimes SO(9)}}$ covariant  Cartan forms (where generalization means the use of $SO(1,10)$ Lorentz covariant derivative instead of the usual derivatives in the definition of proper Cartan forms). In the case of M$0$-brane, which is massless superparticle, the form $\Omega^{\# i}$ is actually not covariant under the whole set of bosonic gauge symmetries (SO(1,10)$\otimes$ $\{$[SO(1,1)$\otimes$SO(9)]$\subset\!\!\!\!\!\!\times$ $K_9\}$ in the case of curved target superspace) but transforms as connections under $K_9$ transformations (see footnote \ref{footK9}). In contrast, $\Omega^{=i}$ is covariant and generalizes the ${SO(1,10)\over {[SO(1,1)\otimes SO(9)]\subset\!\!\!\!\times K_9}}$ Cartan form.

The forms $\Omega^{\# i}$, $\Omega^{=i}$ obey the generalized Peterson-Codazzi equations (see \cite{bpstv,IB+AN=CQG97})
\begin{eqnarray}\label{M0:DOm=} D\Omega^{= i} =  \hat{R}{}^{= i}\; , \qquad  D\Omega^{\# i} =  \hat{R}{}^{\# i}\;  \qquad  (\, \hat{R}{}^{\# i}:=  \hat{R}^{ {a} {b}}u_{ {a}}^{\# } u_{ {b}}{}^i \; ,  \quad \hat{R}{}^{ =i}:=  \hat{R}^{ {a} {b}}u_{ {a}}^{=} u_{ {b}}{}^i \,) \; .  \qquad
\end{eqnarray}
The induced Riemann curvature 2-form ($d\Omega^{(0)}$ from  $DDu_{ {a}}^{\# (=)} = ^{_{\;\, +}}_{^{(-)}}2 d\Omega^{(0)}u_{ {a}}^{\# (=)} + \hat{R}_{ {a}}{}^{ {b}}u_{ {b}}^{\# (=)}$) and the curvature of normal bundle of the worldline superspace (${\cal G}^{ij}$ from  $DDu_{ {a}}^{i} = u_{ {a}}^{j} {\cal G}^{ji} + \hat{R}_{ {a}}{}^{ {b}}u_{ {b}}{}^{i}$)
are defined by the Gauss and Ricci equations
\begin{eqnarray}\label{M0:Gauss}
 d\Omega^{(0)} &= & {1\over 4 } \hat{R}{}^{=\; \# }+  {1\over 4 } \Omega^{=\, i} \wedge \Omega^{\# \, i}\; , \qquad \hat{R}{}^{=\; \# }:=  \hat{R}^{ {a} {b}}u_{ {a}}^{=} u_{ {b}}^{\# } \; , \qquad \\
\label{M0:Ricci}
{\cal G}^{ij} & =& \hat{R}{}^{ij}-   \Omega^{=\, [i} \wedge \Omega^{\# \, j]}\; , \qquad
\hat{R}{}^{ij}:=  \hat{R}^{ {a} {b}}u_{ {a}}^{i} u_{ {b}}{}^j  \; .  \qquad
\end{eqnarray}

Using Eqs. (\ref{M0:Du=Om}), one can specify the selfconsistency (integrability) conditions for the superembedding equation (\ref{Eua=e++u--}) as follows
\begin{eqnarray}\label{T=De++u--} \hat{T}{}^{ {a}}:= - i
{\hat{\cal E}} \wedge \Gamma^{ {a}}{\hat{\cal E}} = {1\over 2} De^{\# }\, u^{ {a}=} + {1\over 2} \, e^{\# } \wedge \Omega^{= i} \; .  \qquad
\end{eqnarray}

\subsubsection{Fermionic conventional constraints and spinor moving frame variables}

To move further, we need to define the fermionic supervielbein of the worldline superspace ${\cal W}^{(1|16)}$. It is convenient to do this by specifying the pull--back of the target superspace fermionic supervielbein  form to ${\cal W}^{(1|16)}$
\begin{eqnarray}\label{Ef=efv+}
\hat{{\cal E}}^{ {\alpha}}=  e^{+q} v_{q}^{- {\alpha}} + e^{\# }\chi_{\# }{}^{-}_q v_{q}^{+ {\alpha}}\; .
 \qquad
\end{eqnarray}
This expression involves so--called spinor moving frame variables or spinorial harmonics (see
\cite{bpstv,IB+AN=CQG97,IB07:M0} and refs. therein). These can be considered as $32\times 16$ blocks of the $Spin(1,10)$ valued moving frame matrix
\begin{eqnarray}\label{Vharm-=M0}
 V_{( {\beta})}{}^{ {\alpha}}= \left(
v_{q}^{- {\alpha}},~ v_q^{+ {\alpha}} \right)\;
\in \; Spin(1,10) \; , \qquad
\end{eqnarray}
and are related to the moving frame vectors by the equations
\begin{eqnarray}\label{M0:v+v+=u++}
 v_{q}^+ {\Gamma}_{ {a}} v_{p}^+ = \; u_{ {a}}{}^{\# } \delta_{qp}\; , \qquad & 2 v_{q}^{+ {\alpha}}v_{q}^{+}{}^{ {\beta}}= \tilde{\Gamma}^{ {a} {\alpha} {\beta}} u_{ {a}}{}^{\# }\; , \qquad
 \\ \label{M0:v-v-=u--}
 v_{q}^- {\Gamma}_{ {a}} v_{p}^- = \; u_{ {a}}{}^{=} \delta_{qp}\; , \qquad & 2 v_{q}^{- {\alpha}}v_{q}^{-}{}^{ {\beta}}= \tilde{\Gamma}^{ {a} {\alpha} {\beta}} u_{ {a}}{}^{=}\; , \qquad
 \\  \label{M0:v-v+=ui}
 v_{q}^- {\Gamma}_{ {a}} v_{p}^+ = - u_{ {a}}{}^{i} \gamma^i_{qp}\; , \qquad & 2 v_{q}^{-( {\alpha}}\gamma^i_{qp}v_{p}^{+}{}^{ {\beta})}=-  \tilde{\Gamma}^{ {a} {\alpha} {\beta}} u_{ {a}}{}^{i}\; , \qquad
\end{eqnarray}
where $\gamma^i_{qp}=\gamma^i_{pq}$ are the $SO(9)$ Dirac matrices. Notice that d=9 charge conjugation matrix is symmetric which allowed to choose it equal to $\delta_{qp}$ and to do not distinguish the upper and lower case indices of $Spin(9)$.

Eqs. (\ref{M0:v+v+=u++})--(\ref{M0:v-v+=ui}) can be obtained  from the condition of the Dirac matrix conservation under Lorenz rotations described by moving frame superfields,
$V {\Gamma}_{ {a}}V^T= u_{ {a}}^{( {b})}{\Gamma}_{( {b})}$ and
$V^T \tilde{\Gamma}^{( {a})}V= \tilde{\Gamma}{}^{ {b}} u_{ {b}}^{ ( {a})}$, by using  a suitable $SO(1,1)\otimes SO(9)$ invariant representation for the Dirac and charge conjugation matrices. These equations, together with the condition of the charge conjugation matrix preservation under Lorentz rotation described by the moving frame variables, $VCV^T=C$, are the constraints which define the spinor moving frame variables. The latter implies that the elements of the inverse moving frame matrix,
\begin{eqnarray}\label{Vharm=M0}
 V^{( {\beta})}_{ {\alpha}}= \left(
v_{ {\alpha}q}{}^+\, ,v_{ {\alpha}q}{}^- \right)\;
\in \; Spin(1,10) \; , \qquad \cases{ v_{q}^{- {\alpha}}v_{ {\alpha}p}{}^+=\delta_{qp}= v_{q}^{+ {\alpha}}v_{ {\alpha}p}{}^-\; , \cr  v_{q}^{- {\alpha}}v_{ {\alpha}p}{}^-= \; 0\; =\; v_{q}^{+ {\alpha}}v_{ {\alpha}p}{}^+\; , }
\end{eqnarray}
are related with $v_{q}^{+ {\alpha}}$ by
\begin{eqnarray}\label{v-1=iCv}
v_{q}^{\pm {\alpha}}= \pm i C^{ {\alpha} {\beta}} v_{ {\beta}q}{}^{\pm} \; . \qquad \end{eqnarray}
The $SO(1,10)\otimes SO(1,1)\otimes SO(9)$ covariant derivatives of the moving frame variables are expressed through  the generalized Cartan forms of Eqs. (\ref{M0:Du=Om}) by the following equations
(see \cite{IB07:M0} and refs. therein for the details of  derivation)
\begin{eqnarray}
\label{Dv-q} &  Dv_q^{-\alpha}  = - {1\over 2} \Omega^{=i}
v_p^{+\alpha} \gamma_{pq}^{i}\; , \qquad \\
\label{Dv+q} &  Dv_q^{+\alpha} = - {1\over 2} \Omega^{\# i}
v_p^{-\alpha} \gamma_{pq}^{i}\; . \qquad
\end{eqnarray}

Using (\ref{Vharm=M0}) we can split Eq. (\ref{Ef=efv+}) on two equations
\begin{eqnarray}\label{ef=}
 e^{+q}= \hat{{\cal E}}^{+q}&:=& \hat{{\cal E}}^{ {\alpha}} v_{ {\alpha}q}{}^+
 \; , \qquad \\ \label{Ef-q=}
\hat{{\cal E}}^{-q}&:=&\hat{{\cal E}}^{ {\alpha}} v_{ {\alpha}q}{}^-
 =  e^{\# }\chi_{\# }{}^{-}_q \; ,
 \qquad
\end{eqnarray}
the first of which manifests the conventional constraint defining fermionic supervielbein of the worldline superspace ${\cal W}^{(1|16)}$ to be induced by the superembedding.

The second equation, (\ref{Ef-q=}), states that $v_{ {\alpha}q}{}^-$ projection of  pull--back to ${\cal W}^{(1|16)}$ of the target superspace fermionic supervielbein, $\hat{{\cal E}}^{-q}$, has no projection on the fermionic supervielbein. One can start from the most general decomposition $\hat{{\cal E}}^{-q}:=\hat{{\cal E}}^{ {\alpha}} v_{ {\alpha}q}{}^-
 =  e^{+p}h^{=}_{pq}+  e^{\# }\chi_{\# }{}^{-}_q$. However then, checking the selfconsistency condition (\ref{T=De++u--}) for the superembedding equation, one arrives (in its {\it dim 1}, $\propto e^{+q}\wedge e^{+p} u_{ {a}}{}^{\# }$ component) at $h^{=}(h^{=})^T=0$. In the case of real number valued matrix elements the only solution of this equation is $h^{=}_{qp}=0$. In the supersymmetric case one can consider a nonzero solution but with all the matrix elements being nilpotent. The r\^ole of such a solution and of its nilpotent Grassmann even parameters is presently unclear so that below we restrict ourselves by considering the trivial solution $h^{=}_{qp}=0$ only. Moreover, we have allowed ourselves a shortcut and have used this solution in Eqs. (\ref{Ef=efv+}) and (\ref{Ef-q=}), specifying the pull--back of the target space fermionic supervielbein form, from the very beginning.

\subsection{Consequences of superembedding equations.  M$0$--brane equations of motion from superembedding}

Substituting Eq. (\ref{Ef=efv+}) into the integrability conditions for superembedding equation, Eq. (\ref{T=De++u--}), one finds, besides  the expression for the worldvolume bosonic torsion 2-form,
\begin{eqnarray}
\label{De++=efef} De^{\# }=-2ie^{+ q} \wedge e^{+q}\; , \qquad
\end{eqnarray}
also the expression for dim 1/2 ($\propto e^{+q}$) component of the generalized Cartan form
$\Omega^{= i}$ entering Eqs. (\ref{M0:Du=Om}) and (\ref{Dv-q}), $\Omega^{= i}_{+q}=-4i\gamma^i_{qp}\chi_{\# p}{}^-$. This implies that
\begin{eqnarray}
\label{Om--i=} \Omega^{= i}_{+q}=-4ie^{+q}\gamma^i_{qp}\chi_{\# p}{}^- + e^{\# } \Omega_{\# }^{\; = i}\; , \qquad \end{eqnarray}
where
\begin{eqnarray}\label{chi=Om=}
\chi_{\# p}{}^-:= \hat{{\cal E}}_{\# }{}^{ {\alpha}}v_{ {\alpha}p}{}^- \; , \qquad \Omega_{\# }^{\; = i}:= D_{\# } u^{=  {a}}\, u_{ {a}}{}^i = D_{\# } \hat{E}_{\# }{}^{  {a}}\, u_{ {a}}{}^i\; . \qquad
\end{eqnarray}
Eqs. (\ref{chi=Om=}), which resume some consequences of Eqs. (\ref{Ef=efv+}), (\ref{M0:Du=Om}) and (\ref{Eua=e++u--}), show that the bosonic and fermionic equations of motion for M$0$--brane can be formulated in terms of the differential form $\Omega^{= i}$ (see \cite{IB+AN=CQG97} for similar statements in $D=10$ superparticle case).

At this stage it is useful to calculate the pull--back of the fermionic torsion (\ref{Tf=}),
\begin{eqnarray} \label{M0:hTf=}
\hat{T}{}^{ {\alpha}} =- {1 \over 72} e^{\# } \wedge e^{+q} \hat{F}^{= ijk}\gamma^{ijk}_{qp} v_p^{- {\alpha}} \; , \qquad \hat{F}^{= ijk}:= F^{ {a} {b} {c} {d}} (\hat{Z}) u_{ {a}}
{}^{ =}u_{ {b}}{}^{i}u_{ {c}}{}^{j}u_{ {d}}{}^{k}\; ,
\end{eqnarray}
and substitute  it in the integrability conditions for the fermionic differential form equation  (\ref{Ef=efv+}),
\begin{eqnarray} \label{D(Ef-ev-)=0}D(\hat{{\cal E}}^{ {\alpha}}- e^{+q} v_{q}^{- {\alpha}}- e^{\# }\chi_{\# }{}^{-}_q v_{q}^{+ {\alpha}})=0\; . \qquad
\end{eqnarray} The lowest dimensional (dim 1) component of the $v_{ {\alpha}q}{}^-$ projection of this equation results in requirement that $(\chi_{\# }{}^-\gamma^i)_{(r} \gamma^i_{p)q} = \delta_{pq} \chi_{\# q}{}^-$. This equation has only trivial solution, which implies that the superembedding equation results in the fermionic equations of motion for the M$0$--brane, $\chi_{\# }{}^{-}_q=0$. The next, dim 3/2 component of the $v_{ {\alpha}q}{}^-$ projection  of Eq. (\ref{D(Ef-ev-)=0}) results in $D_{+p}\chi_{\# q}{}^-= -{1\over 2}\gamma^i_{pq} \Omega_{\# }^{\; -- i}$, so that the above fermionic equation implies the bosonic equation of motion $\Omega_{\# }^{\; -- i}=0$. To resume, we have shown how the superembedding equation leads to the dynamical bosonic and fermionic equations of motion of the M$0$--brane,
\begin{eqnarray}\label{M0:DiracEq}
\chi_{\# p}{}^-&:=& \hat{{\cal E}}_{\# }{}^{ {\alpha}}v_{ {\alpha}p}{}^- =0 \; , \qquad  \\ \label{M0:bEq} \Omega_{\# }^{\; = i}\, &:=& D_{\# } u^{=  {a}}\, u_{ {a}}{}^i = D_{\# } \hat{E}_{\# }{}^{  {a}}\, u_{ {a}}{}^i=0 \; . \qquad
\end{eqnarray}
Let us stress that the above relations are equations of motion for M$0$-brane in an arbitrary $D=11$ supergravity background. When passing from flat to curved target superspace, no explicit fluxes appear in the right--hand side of these equations ({\it cf.} $D=10$ D$0$-brane case \cite{IB09:D0}), but the bosonic and fermionic supervielbein, the pull--back of which to ${\cal W}^{(1|16)}$ enter  (\ref{M0:DiracEq}), acquire the contribution form the supergravity fields.

Notice that, besides the equations of motion, the investigation of Eq. (\ref{D(Ef-ev-)=0}) (of its $v_{ {\alpha}q}{}^+$ projection) allows to find the fermionic torsion of the worldvolume geometry induced by superembedding,
\begin{eqnarray} \label{M0:De+q=}
De^{+q}=- {1 \over 72} e^{\# } \wedge e^{+q} \hat{F}^{= ijk}\gamma^{ijk}_{qp}  \; , \qquad \hat{F}^{= ijk}:= F^{ {a} {b} {c} {d}} (\hat{Z}) u_{ {a}}
{}^{ =}u_{ {b}}{}^{i}u_{ {c}}{}^{j}u_{ {d}}{}^{k}\; .
\end{eqnarray}
The influence of the  4-form flux on  the induced geometry of ${\cal W}^{(1|16)}$ is seen in this equation and also in  the expression for the curvature of normal bundle (see (\ref{M0:Ricci}),
\begin{eqnarray} \label{M0:Gij=onshell}
{\cal G}^{ij}= \hat{R}^{ij}= e^{+q}\wedge e^{+p} \left( {2i\over 3}\hat{F}{}^{=\, ijk}\gamma^{k}_{qp}+ {i\over 18}\hat{F}{}^{=\, klm}\gamma^{ijklm}_{qp} \right) + i  e^{\# }\wedge e^{+q} \hat{T}^{=\, [i}{}^-_p \gamma^{j]}_{pq}    \;  \qquad
\end{eqnarray}
($\gamma^{i_1\ldots i_5}_{qp}= {1\over 4!} \epsilon^{i_1\ldots i_5j_1\ldots j_4}_{qp} \gamma^{j_1\ldots j_4}_{qp}\,$, $\; \hat{T}^{=\, i}{}^-_q :=\hat{T}_{ {a} {b}}{}^{ {\beta}}u^{a=}u^{ {b}i}v_{ {\beta}q}{}^-$).  The Riemann curvature of ${\cal W}^{(1|16)}$ (see (\ref{M0:Gauss})) vanishes on the mass shell, $d\Omega^{(0)}=0$.

\section{Searching for multiple M$0$-equations in the frame of superembedding approach}

In this section we will report first results of the search  for multiple M$0$-brane equations in the frame of superembedding approach, which can be considered as the search for restoration of $D=11$ Lorentz invariance in the multiple D$0$-brane equations of \cite{IB09:D0}. As far as the latter are obtained by considering the constraints of maximally supersymmetric $SU(N)$ YM gauge theory on the worldline superspace of D$0$-brane, it is natural to search for multiple M$0$--brane equations by studying the possible constraints imposed on the superfield strength
\begin{eqnarray}\label{M0:G=dA-=}
G_2= dA - A\wedge A= {1\over 2}e^{+q}\wedge e^{+p}
G^{--}_{q\, p} + e^{\# }\wedge e^{+q} G_{+q\, \# }\;  \qquad
\end{eqnarray}
of the $su(N)$-valued  1-form gauge potential $A=e^{\# }A_{\# }+ e^{+q} A_{+q}$ defined on $d=1$ ${\cal N}=16$ superspace ${\cal W}^{(1|16)}$. It is also natural to assume that the embedding of ${\cal W}^{(1|16)}$  into the 11D target superspace is determined by superembedding equation (\ref{Eua=e++u--}),
 \begin{eqnarray}
\label{mM0:Eua=e++u--}
  \hat{E}^{ {a}} = {1\over 2} e^{\# } u^{ {a}=} \;
\qquad
\end{eqnarray}
(see \cite{IB09:D0} for more discussion). Then this superspace describing the `center of mass'  motion of the multiple M$0$--system is described by the superfield equations of  Sec. 2.

We have put quotation mark on `center of mass' as M$0$-branes are massless superparticles, so that one should rather say  'center of energy'.  As the superembedding equation for ${\cal W}^{(1|16)}\subset \Sigma^{(11|32)}$ puts the theory on mass shell, one finds that this center of energy motion is characterized by light-like geodesic in the bosonic body of the target 11D superspace.
As an analogy, one can imagine the system of multiple M$0$-s as  the beam of light, the trajectory of which is light-like, but with an interaction between the (originally massless) constituents which is assumed to be described by the above $d=1$ ${\cal N}=16$ $SU(N)$ gauge supermultiplet.

\subsection{Constraints on the field strength of $d=1$ ${\cal N}=16$ $SU(N)$ SYM on ${\cal W}^{(1|6)}$}

The natural candidate for the constraints, which is also suggested by the D$0$-brane description of \cite{IB09:D0}, reads\footnote{Although below we do not specify the $SO(1,1)$ weights of the matrix superfields, it is convenient to remember that, {\it e.g.}  ${\bb X}{}^i:= {\bb X}_{\# }{}^i\equiv  {\bb X}_{++ }{}^i\equiv {\bb X}{}^{--i}=:{\bb X}{}^{=i}$. }
\begin{eqnarray}\label{M0:G=sX}
G^{--}_{q\, p}= i \gamma^i_{qp} {\bb X}{}^i \; . \qquad
\end{eqnarray}
The nanoplet of $su(N)$-valued superfields ${\bb X}{}^i$ ($={\bb X}_{\# }{}^i$) provides a natural candidate for the description of the relative motion of the M$0$--brane constituents. The lowest dimensional (dim 1, $\propto e^{+q}\wedge e^{+p}$) component of the Bianchi identities $DG=0$ requires  that ${\bb X}{}^i$  obeys the following superembedding--like superfield  equation\footnote{To arrive at this conclusion, one uses the $d=9$ $\gamma$--matrix identity $\gamma^{i}_{(pq}\gamma^{i}_{r)s}= \delta_{(pq}\delta_{r)s}$. Among their other useful properties are $\delta_{p(r}\delta_{s)q} = {1\over 16}\delta_{pq}\delta_{rs} + {1\over 16}\gamma^{i}_{pq}\gamma^{i}_{rs}+ {1\over 16\cdot 4!}\gamma^{ijkl}_{pq}\gamma^{ijkl}_{rs} $  and $\gamma^{ijkl}_{(pq}\gamma^{ijkl}_{r)s}= 14\cdot 4!\delta_{(pq}\delta_{r)s}$.}
\begin{eqnarray}\label{M0:DX=gP}
D_{+q}{\bb X}{}^i= 4i\gamma^i_{qp}\Psi_{q}\; , \qquad
\end{eqnarray}
and identify the fermionic $su(N)$ valued (Hermitian traceless) matrix superfield  $\Psi_{q}$  with the dim 3/2 field strength in (\ref{M0:G=dA-=}),
\begin{eqnarray}\label{M0:G=P}
G_{+q\, \# }= -i\Psi_{q}\; . \qquad
\end{eqnarray}
The highest dimensional (dim 3/2, $\propto e^{\# }\wedge e^{+q}$) component of this Bianchi identity is satisfied identically when the consequences of the  integrability conditions for the superembedding--like equation (\ref{M0:DX=gP}) are taken into account. These include $8i\gamma^i_{r(p}D_{+q)}\Psi_{r}= 4i \delta_{qp}D_{\# }{\bb X}^i + i \gamma^j_{qp}[{\bb X}^j, {\bb X}^i] + {\bb X}^j {\cal G}_{+q+p}^{ji}$ which, after some algebra,
results in
\begin{eqnarray}\label{M0:Dpsiq=}
D_{+q}\Psi_{p}\; = {1\over 2}\gamma^i_{qp}D_{\# }{\bb X}^i & +& {1\over 16}
\gamma^{ij}_{qp} \; \left([{\bb X}^i, {\bb X}^j]
-{4\over 3}  \hat{F}{}^{= ijk}{\bb X}^k \right) -  {1\over 72} \gamma^{ijkl}_{qp} {\bb X}^i
\hat{F}{}^{= jkl}  \; . \qquad
\end{eqnarray}

\subsection{Equations of motion for multiple M$0$-brane system in flat target $D=11$ superspace}

For simplicity, we restrict our discussion below by the case of multiple M$0$-brane system in flat target superspace,  where, in particular, $\hat{F}{}^{= jkl}=0$, $De^{+q}=0$, ${\cal G}^{ij}=0$ and  Eq. (\ref{M0:Dpsiq=}) simplifies to  \begin{eqnarray}\label{M0:Dpsiq=fl}
D_{+q}\Psi_{p}\; = {1\over 2}\gamma^i_{qp}D_{\# }{\bb X}^i + {1\over 16}
\gamma^{ij}_{qp} \; [{\bb X}^i, {\bb X}^j]
 \; . \qquad
\end{eqnarray}
The investigation of the more general case of curved supergravity superspace, including the study of the dielectric brane effect for the multiple M$0$--system, will be the subject of the separate study \cite{mM0+flux}.

Now, the integrability condition for Eq. (\ref{M0:Dpsiq=fl}) ($D_{+(p}D_{+q)}\Psi_r=...$)  reads
\begin{eqnarray}\label{M0:dD++Psi++=}
& (\delta_{qp}\delta_{rs}- \gamma^i_{r(p}\gamma^i_{q)s}) D_{\# }\Psi_{q}=
{1\over 4}\left(\gamma^i_{r(p}\delta_{q)s}- \gamma^i_{pq}\gamma^i_{rs}-\gamma^{ij}_{r(p}\gamma^j_{q)s}\right) \left[{\bb X}^i\, , \, \Psi_s\right]\; . \qquad
\end{eqnarray}
The $\delta_{qp}$ trace part of (\ref{M0:dD++Psi++=}) gives the fermionic equations of motion
\begin{eqnarray}\label{M0:D++Psi++=}
& D_{\# }\Psi_q+
{1\over 4} \gamma^i_{qp} \left[ {\bb X}^i\, , \, \Psi_p \right]=0\; , \qquad
\end{eqnarray}
while the other irreducible parts are satisfied identically after Eq. (\ref{M0:D++Psi++=}) is taken into account. Acting on Eq. (\ref{M0:D++Psi++=}) by Grassmann covariant derivative $D_{+p}$ and decomposing the result on irreducible representations of $SO(9)$ one finds the bosonic equation of motion
\begin{eqnarray}\label{M0:DDX++i=}
& D_{\# }D_{\# } {\bb X}^i =
{1\over 16} \left[ {\bb X}^j\, , \, \left[ {\bb X}^j\, , \, {\bb X}^i\,  \right]\right]+ i\gamma^i_{qp} \left\{\Psi_q\, , \, \Psi_p \right\}\; , \qquad
\end{eqnarray}
as well as the constraint
\begin{eqnarray}\label{M0:[XDX]=}
& \left[ {\bb X}^i\, , \, D_{\# } {\bb X}^i \right]= - 4i \left\{\Psi_q\, , \, \Psi_q \right\}\; , \qquad
\end{eqnarray}
This latter is the counterpart of the  Gauss constraint of the multiple D$0$-brane system obtained from the superembedding description  in \cite{IB09:D0}.

Thus our superembedding description is consistent and produces the dynamical equations for the multiple M$0$-brane system in flat target superspace given in  Eqs. (\ref{M0:DiracEq}), (\ref{M0:bEq}),  (\ref{M0:D++Psi++=}) (\ref{M0:DDX++i=}) and (\ref{M0:[XDX]=}). Our approach can be considered as a covariant generalization of the Matrix model. The study of the system of equations for multiple M$0$-system in the nontrivial $D=11$ supergravity background, including the description of dielectric brane effect in this case, will be discussed elsewhere \cite{mM0+flux}.


\section{Conclusion and discussions}
\label{Conclusion}
In this letter we have studied a possibility to describe multiple M$0$-brane system in the frame of superembedding approach.  One of the motivation for such a study is that existence of such a system can be considered as an indication of restoration of the $11D$ Lorentz symmetry in the superfield description of multiple D$0$--brane system of  \cite{IB09:D0}. It is also of interest in the context of searching for a  complete, covariant and supersymmetric generalization of the Matrix model equations.

We have studied  the simplest version of the superembedding approach based on the most natural 1d ${\cal N}=16$  SYM constraints imposed on the $su(N)$ valued 2-superform field strength defined on superspace  ${\cal W}^{(1|16)}$ with one bosonic and $16$ fermionic directions.  The embedding of ${\cal W}^{(1|16)}$ into the target $11D$ superspace ${\Sigma}^{(11|32)}$ obeys superembedding equation characteristic for a single M$0$-brane. We have shown that such a description is consistent and produces the interacting bosonic and fermionic equations of motion for the (super)field describing relative motion of the constituents of multiple M$0$--system. Although the basic equations of our formalism and some of their consequences have been written in general $D=11$ supergravity background, the equations of motion have been presented for the case of multiple M$0$-brane system in flat target superspace only. The detailed study of the multiple M$0$ equations in the general curved 11D supergravity superspace, which shall provide a covariant generalization of Matrix model equation to the case of nontrivial supergravity background, will be the subject of future paper \cite{mM0+flux}.

\subsection*{Acknowledgments}

The author thanks Djordje Minic and Martin Kruczenski for useful conversations as well as  Thomas  Curtright, Jo Ann Curtright and Luca Mezincescu for the hospitality at the Miami 09 Conference in Fort Lauderdale, where this work was essentially completed. The partial support  by the research grants FIS2008-1980 from the Spanish MICINN  and 38/50--2008 from Ukrainian National Academy of Sciences and Russian Federation RFFI are greatly acknowledged.

\subsection*{Notice added}

After the first version of this letter had appeared on the net the author became aware about the paper \cite{YLozano+=0207} where a counterpart of the Myers action for the case of multiple M0-brane (`multiple M-theory gravitational waves/gravitons') was proposed and studied. This (purely bosonic) action was obtained starting from the counterpart of the Myers action for multiple type IIA gravitons  \cite{YLozano+=0205}.

 \end{document}